\documentstyle[epsfig]{aipproc}
\newcommand{\fbfive}{F(162){}}
\newcommand{\mbfive}{m(162){}}

\newcommand{\ha}{H$\alpha${}}

\newcommand{\msol}{M$_\odot$}

\newcommand{\hi}{H{\small I}}
\newcommand{\hii}{H{\small II}}

\newcommand{\apj}{{ApJ}}
\newcommand{\apjs}{{ApJ Supp.}}
\newcommand{\aj}{{AJ}}
\newcommand{\aaps}{{A\&A Supp.}}

\newcommand{\mnras}{{MNRAS}}

\begin{document}
\title{UIT Observations of the SMC}

\author{Robert H. Cornett$^*$, Theodore P. Stecher \\
and the UIT Science Team} 
\address{Code 681, LASP/GSFC, Greenbelt MD 20771\\
$^*$ Hughes STX Corp.}

\maketitle

\begin{abstract}

A mosaic of four UIT far-UV (FUV; 1620\AA) images, which covers most of the 
SMC bar, is presented, with derived stellar and HII region photometry. 
The UV morphology of the SMC's Bar shows that recent star formation there 
has left striking features including: a) four concentrations of UV-bright 
stars spread from northeast to southwest at nearly equal 
($\sim$30 arcmin=0.5 kpc) spacings; b) one concentration
comprising a well-defined 8-arcmin diameter ring surrounded by a 
larger \ha\ ring, suggestive of sequential star formation.

FUV PSF photometry is obtained for 11,306 stars, and FUV photometry is obtained
for 42 \ha-selected \hii\ regions, both for the stars and for the total
emission contained in the apertures defined by KH \cite{kh}.
The  flux-weighted average ratio of total to stellar FUV flux
is 2.15; the stellar FUV luminosity function indicates
that most of the excess total flux is due to scattered FUV radiation, rather
than faint stars.  Both stellar and total emission are well
correlated with \ha\ fluxes measured by KH, and yield FUV/\ha\ flux ratios
that are consistent with models of single-burst clusters with SMC metallicity,
ages from 1-5 Myr, and moderate (E(B-V)=0.0-0.1 mag) internal SMC extinction.

\end{abstract}

\begin{figure}[b!] 
\centerline{\epsfig{file=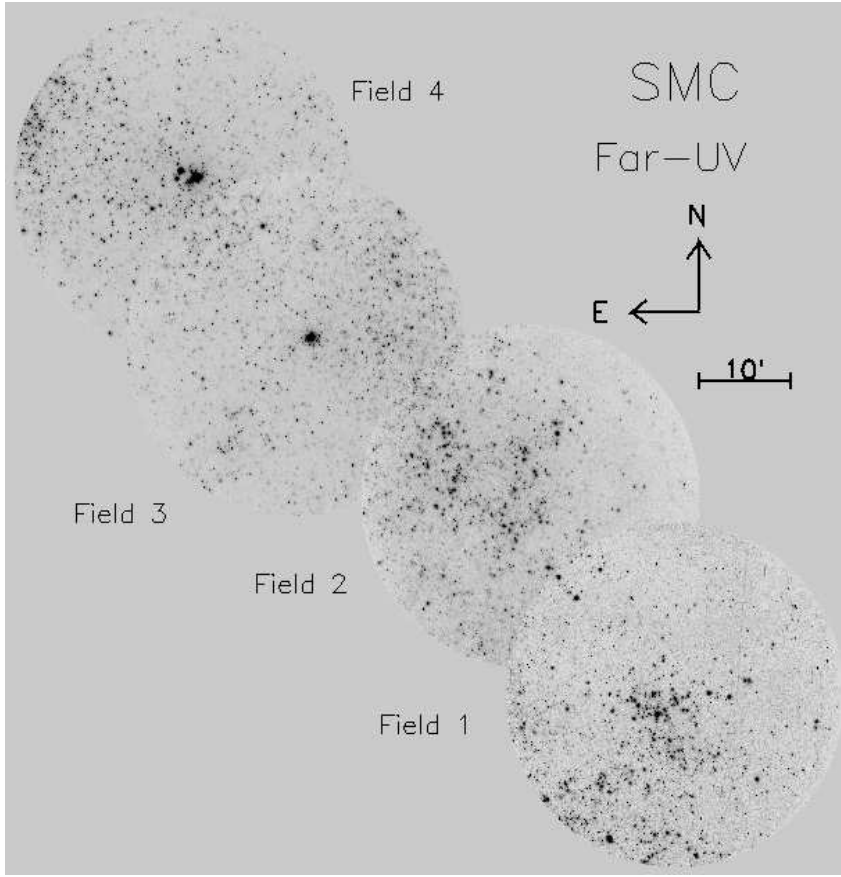,height=4.5in,width=4.5in}}
\caption{FUV Mosaic of the SMC Bar} 
\label{cornettmosaic}
\end{figure}

UV observations from above the earth's atmosphere
are vital for understanding Population I properties of metal-poor, 
``primitive''galaxies such as the Small Magellanic Cloud 
(SMC) \cite{cornetta}.  Many effects of composition differences appear 
best, or only, in the UV. Line blanketing strongly affects UV colors; the steep
SMC extinction curve (A$_{162}/E(B-V)\sim16$), widely thought to be due
to abundances in SMC dust, is ``extreme'' only in the UV; and FUV
photometry is more effective than optical-band photometry in determining 
temperatures of hot stars.  Here, we present initial results based on a 
mosaic of four, 40-arcmin diameter, 3-arcsec-resolution, far-UV images, 
nearly covering the SMC bar, (figure \ref{cornettmosaic}) obtained by the 
Ultraviolet Imaging Telescope (UIT) during the Astro missions. 
Details of UIT hardware, calibration, operations, and data reduction 
are in \cite{stechera}, and detailed results of the current study are 
in \cite{cornettb}.

Figure \ref{cornettmosaic} shows that the SMC's FUV emission originates 
mostly in hot stellar populations which, while not restricted to clusters,
are significantly clumped.  No diffuse FUV emission is readily apparent. 
The brightest features are NGC 346 and NGC 330, in Fields 4 and 3,
with additional FUV concentrations centered in fields 1 and 2.
The bright FUV concentrations, spaced along the Bar at
$\sim$0.5kpc intervals, have similar clustering and distribution properties to
those evident in wide-field FUV images of the LMC \cite{asmith}.

UIT field 2, near the Bar's center, provides an intriguing instance of what
appears to be sequential star formation.  A ring of FUV-bright stars 
dominates the field, and other evidence, including 
\hi\ shells  \cite{staveleyb} surrounding the stellar ring,
an old supernova remnant 0050-728 \cite{mathe} at its northern edge, and 
broad \ha\ linewidths \cite{okumura} at the ring's center, shows ties 
between the distribution of hot stars and the gas dynamics.  

\begin{figure}[b!] 
\centerline{\epsfig{file=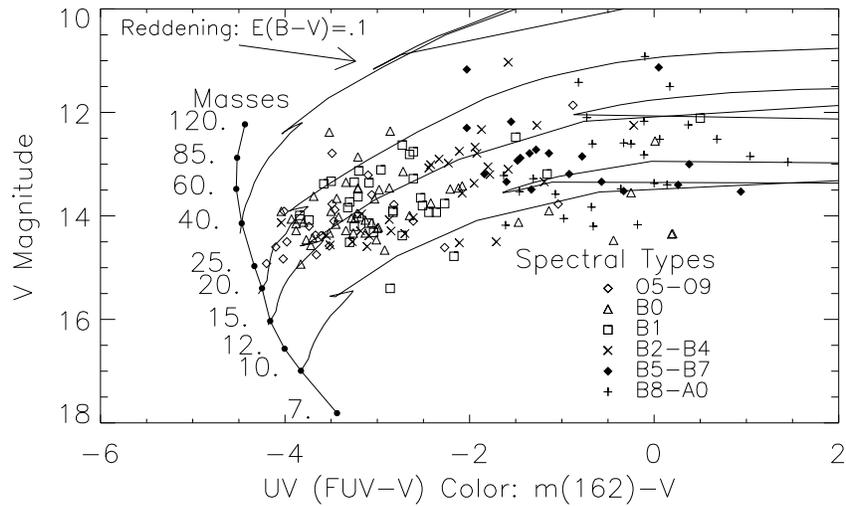,height=3.0in,width=5.in}}
\caption{SMC (FUV-V,V) Color-Magnitude Diagram} 
\label{cornettcmdiag}
\end{figure}


We have derived far-UV PSF photometry for 11,306 stars and correlated 
our observations with ground-based stellar photometry. 
Figure \ref{cornettcmdiag} is the (\mbfive$-$V),V color-magnitude diagram for
191 stars from \cite{azzvig}.  Discrete symbols are observed 
stars, uncorrected for reddening, with spectral types. 
Solid lines and evolutionary tracks show paths of 10, 15, 20, and 40 
\msol\  SMC-composition models (\cite{kurucz}, \cite{char}) corrected for 
distance and foreground Galactic extinction. 
The reddening vector shows  a typical large value for the SMC. The tracks 
show that these stars predominantly have masses 10--20\msol\  and imply ages 
10--30 Myr. The general segregation of spectral types by color implies that 
much of the (\mbfive$-$V) color variation seen in this figure is due to the 
intrinsic colors of the stars themselves.


\begin{figure}[b!] 
\centerline{\epsfig{file=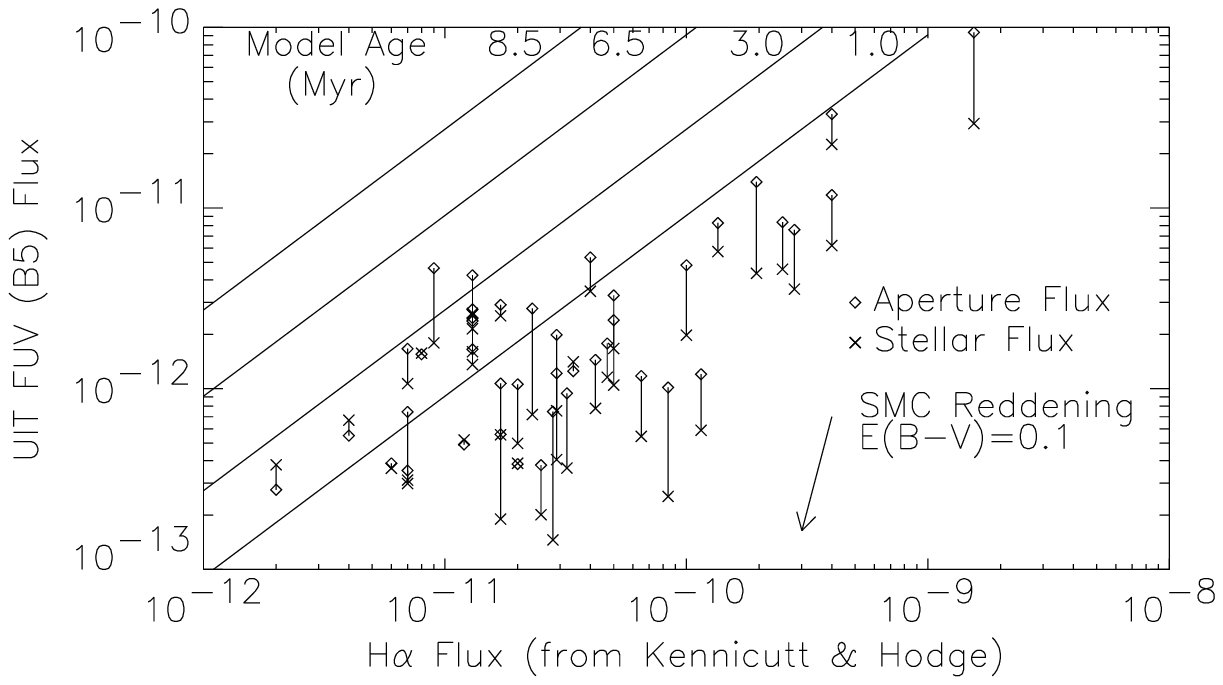,height=3.0in,width=5.in}}
\caption{(FUV vs. \ha\ Flux for \hii\ Regions} 
\label{cornettfluxplot}
\end{figure}

We have measured FUV fluxes for 42 \hii\ regions 
measured at \ha\ by KH \cite{kh}, and compared 
the observed FUV/\ha\ flux ratios with cluster models. 
Figure \ref{cornettfluxplot} shows these data. Open diamonds are
aperture fluxes, and crosses are the sum of stellar fluxes in the 
apertures, uncorrected for Galactic foreground reddening. The FUV flux is 
well correlated with \ha\, and the ratio of FUV aperture
to stellar flux is relatively uniform. The ratio measures 
the relative amounts of ``diffuse'' light (which includes faint stars and 
scattered light) and light from bright stars. 
The flux-weighted average ratio of aperture to stellar flux is 2.15.  
Two arguments point to a dust-scattering origin for most of the 
excess FUV aperture flux, however.  First, the total-to-stellar FUV flux 
ratio for the Orion nebula, with few undetected stars, is 2.5 
\cite{bohlin}.  Second, extrapolating the FUV luminosity function 
predicts an additional stellar flux contribution of only 22\%. In spite of 
low dust abundance, scattering of FUV radiation is important in the 
SMC. 

The FUV/\ha\ ratio is known to be a good diagnostic of \hii\ region 
evolution \cite{rsh}. We have modelled \cite{landsman} \fbfive/\ha\ ratio 
values for SMC-composition clusters with single-burst star formation.
The ratio rises monotonically from 0.1 at 1 Myr to 4.3 at 10 Myr; foreground-
corrected ratios for various cluster ages are
plotted as solid lines in Figure \ref{cornettfluxplot}.  
The observed \hii\ regions are evidently no older than $\sim$5 Myr for any 
internal SMC reddening, and are likely between 
1 Myr and 3 Myr in age.  This result is consistent with observations of \hii\
regions in galaxies as disparate as NGC 4449 \cite{rsh} and
M81 \cite{jkhb}.

\end{document}